\documentclass{ws-procs9x6}

\def\bu{{\mathbf{u}}}

\def\bA{{\mathbf{A}}}
\def\bB{{\mathbf{B}}}
\def\bJ{{\mathbf{J}}}

\newcommand{\ben}{\begin{equation}}
\newcommand{\een}{\end{equation}}
\newcommand{\bea}{\begin{eqnarray}}
\newcommand{\eea}{\end{eqnarray}}
\newcommand{\ba}{\begin{array}}
\newcommand{\ea}{\end{array}}

\def\math{\mathsurround 0pt}
\def\oversim#1#2{\lower.5pt\vbox{\baselineskip0pt \lineskip-.5pt
        \ialign{$\math#1\hfil##\hfil$\crcr#2\crcr{\scriptstyle\sim}\crcr}}}

\begin{document}

\title{Decay of Magnetic Fields in the 
Early Universe\thanks{\textsc{Presented at Strong \& Electroweak Matter 2002, 
Heidelberg, Oct 2--5 2002}}}

\author{MARK HINDMARSH}

\address{
Astronomy Centre and Centre for Theoretical Physics\\
University of Sussex\\
Brighton BN1 9QH\\
U.K.\\
E-mail: m.b.hindmarsh@sussex.ac.uk}

\author{M.~CHRISTENSSON AND A.~BRANDENBURG}

\address{NORDITA\\
Blegdamsveg 17\\
DK-2100 Copenhagen \O\\
Denmark\\
E-mail: mattias@nordita.dk, brandenb@nordita.dk}  

\maketitle

\abstracts{
We study the evolution of a stochastic helical magnetic field generated in the
early Universe after the electroweak phase transition, using standard
magnetohydrodynamics (MHD).  We find how the coherence length $\xi$, magnetic
energy $E_{\rm M}$ and magnetic helicity $H$ evolve with time.  We show that the
self-similarity of the magnetic power spectrum alone implies that $\xi \sim
t^{1/2}$.  This in turn implies that magnetic helicity decays as $H\sim
t^{-2s}$, and that the magnetic energy decays as $E_{\rm M} \sim t^{-0.5-2s}$,
where $s$ is inversely proportional to the magnetic Reynolds number $Re_{\rm M}$.
These laws improve on several previous estimates.\\[2pt]
SUSX-TH-03-002 NORDITA-2003-5 AP
}

\section{Introduction}

Magnetic fields are found everywhere in the Universe, from planets to galaxy
clusters\cite{Zel88,Kro94}.  On the galactic scale and above, the strength is of
order a few $\mu$Gauss, maintained by dynamo action, with a characteristic
timescale of roughly a rotation period, $10^8$ yr.  A seed is required to start
the dynamo, and a simple calculation\cite{Zel88} based on the age of a typical
galaxy shows that the seed field must have been about \(10^{-20}\) Gauss.

There are many ideas for the origin of seed field.  Most conservatively, a
Biermann battery operated at the era of reionisation\cite{GneFerZwe00}.  More
speculatively, a field could have been generated in the early
Universe\cite{Grasso:2000wj}, which would have given rise to stochastic,
homogeneous and isotropic magnetic and velocity fields, characterized by their
power spectra and initial length scales.  In order to know the strength and
coherence scale of this field today, we must study freely decaying
magnetohydrodynamic (MHD) turbulence\cite{BisMul,ChrHinBra01}.  This talk
reports earlier work\cite{Christensson:2002xu} where we have developed a new
framework for understanding decaying 3D MHD turbulence, in the case where the
fields are close to being maximally helical, as in the mechanisms proposed
earlier\cite{HelGen}.

Of particular interest are how the coherence length $\xi$, the magnetic energy
$E_{\rm M}$ and magnetic helicity $H$ evolve with time.  Their growth and decay
are thought to be power laws, resulting from an inverse
cascade\cite{PouFriLeo76}, in which power is transferred locally in $k$-space
from small to large scales.  This is not a new problem:  there are models both
with\cite{Ole97,Son:1999my,Field:2000hi,Shiromizu:1998bc} and 
without\cite{Bis93,Field:2000hi,Son:1999my} helicity, and numerical simulations
have also been performed\cite{BraEnqOle96,BisMul}.
We have now developed\cite{ChrHinBra01,Christensson:2002xu} what we believe is the
definitive answer for the evolution of helical magnetic fields 
between the electroweak phase transition and
the time of $e^+e^-$ annihilation.

\section{3D MHD simulations of decaying turbulence}

The MHD equations in an expanding Universe are most conveniently expressed in
terms of conformally rescaled fields $\bB$ (where $\bB=\nabla\times \bA$ is the
magnetic field in terms of the magnetic vector potential $\bA$), $\bu$ (the
velocity) and dissipation parameters $\nu$ (kinematic viscosity), $\eta$
(magnetic diffusivity)\cite{Subramanian:1998gi,ChrHinBra01}.  The current
density $\bJ = \nabla\times \bB$.

In the ideal limit $\nu=\eta=0$, there is an important conserved quantity in
addition to the total energy, which is the magnetic helicity $H=\int\bA\cdot\bB\,
d^{3}x,$ also known as the Abelian Chern-Simons number.  Also important is the
magnetic Reynolds number $Re_{\rm M} = \xi v/\eta$, where $\xi$ and $v$ are the
typical length scale and velocity of the system, because it measures the
relative size of the non-linear term in the induction equation.

\begin{figure}[ht]
\centering{
\scalebox{0.3}{\includegraphics{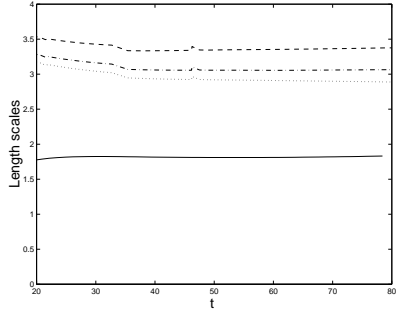}}
}
\caption{Time evolution of the ratio of length scale to the diffusion scale for one run.
The notation is $\xi_{\rm I}/\xi_{\rm diff}$ (dashed),
$\xi_{\rm H}/\xi_{\rm diff}$ (dot-dashed),
$\xi_{\rm R}/\xi_{\rm diff}$ (dotted) and
$\xi_{\rm T}/\xi_{\rm diff}$ (continuous).  The length scales are defined  
in Eq.\  (\ref{length_H}) and the subsequent  
text.
}
\label{length_scales}
\end{figure}

We solve the MHD equations numerically\cite{Bra01}, taking 
$\bu$ and $\bB$ to be homogeneous and isotropic Gaussian random 
fields drawn from a power-law distribution with a high wavenumber cut-off. 
The initial helicity, 
ranged from identically zero to maximal, where the inequality
$
   | H(k)|\leq 2k^{-1} E_{\rm M}(k)
   \label{hel_bound_2}
$
is saturated. Helicity is not exactly conserved in our simulations, and
providing 
that the magnetic energy spectrum $E_{\rm M} (k)$
decays faster than $k^{-2}$, we can define a helicity scale
$\xi_{\rm H}$ such that
\ben
  {\dot H} = -2\eta\,H/{\xi_{\rm H}^{2}}
  \label{length_H}
.\een
If we assume that the evolution of $\xi_{\rm H}$ is described by a power law
$\xi_{\rm H} \sim t^{r}$ it is clear that 
only if $r\,=\,0.5$ does the magnetic helicity show a power law decay
$H=H_0(t/t_0)^{-2s}$
where $s = (\xi_{\rm diff}/\xi_H)^2$, 
in terms of the diffusion scale $\xi_{\rm diff} = 2\pi\sqrt{\eta t}$.
Additional length scales we consider are the integral scale
$\xi_{\rm I} = 2\pi \int dk k^{-1}E_{\rm M}(k) / \int dk E_{\rm M}(k)$, 
the relative helicity scale 
$\xi_{\rm R} = \pi |H| / E_{\rm M}$ and the magnetic Taylor microscale 
$\xi_{\rm T} = 2\pi B_{\rm rms} / J_{\rm rms}$, where $B_{\rm rms}$ and 
$J_{\rm rms}$ are the RMS magnetic field and current density respectively. 
It is plausible that all these scales are proportionally related and 
 Fig.~\ref{length_scales} shows that this is indeed the case.
%******
%******
From the bound on the helicity power spectrum one can show 
$H_{\rm REL} = (\xi_{\rm R}/\xi_{\rm I}) \le 1$, 
so if this bound remains approximately saturated, and the helicity scale goes as  
$\xi_{\rm H} \sim t^{0.5}$, the decay law for
the magnetic energy is
\ben
  E_{\rm M} \sim t^{-0.5 - 2s}.
  \label{energy_dec}
\een
Given   $H_{\rm REL}=\xi_{\rm R}/\xi_{\rm I}$, it is seen from 
Fig.~\ref{length_scales} that $H_{\rm REL}$ is indeed of order unity and 
does not decay markedly with time.

\begin{figure}[ht]
\centering{
\scalebox{0.3}{\includegraphics{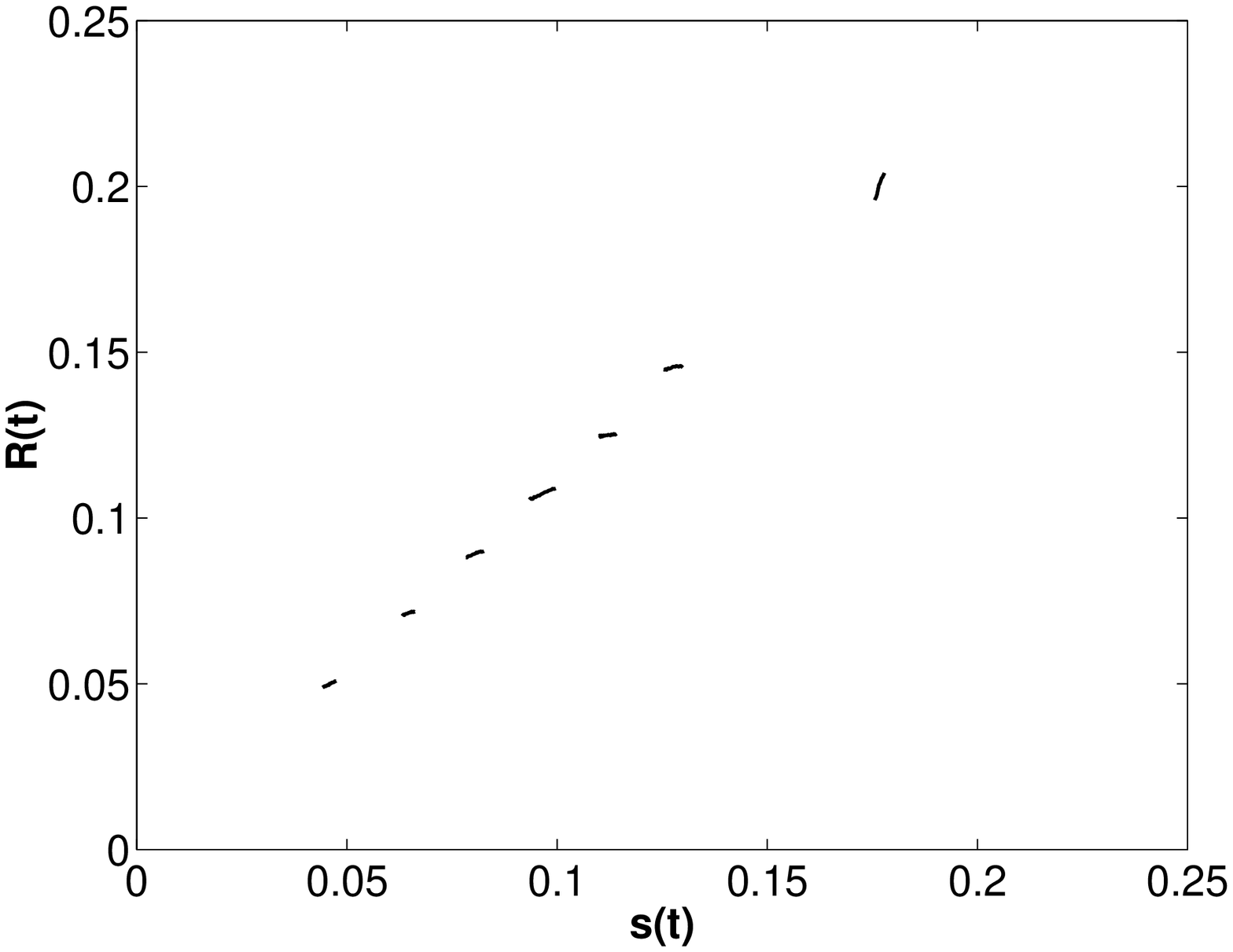}}\scalebox{0.3}{\includegraphics{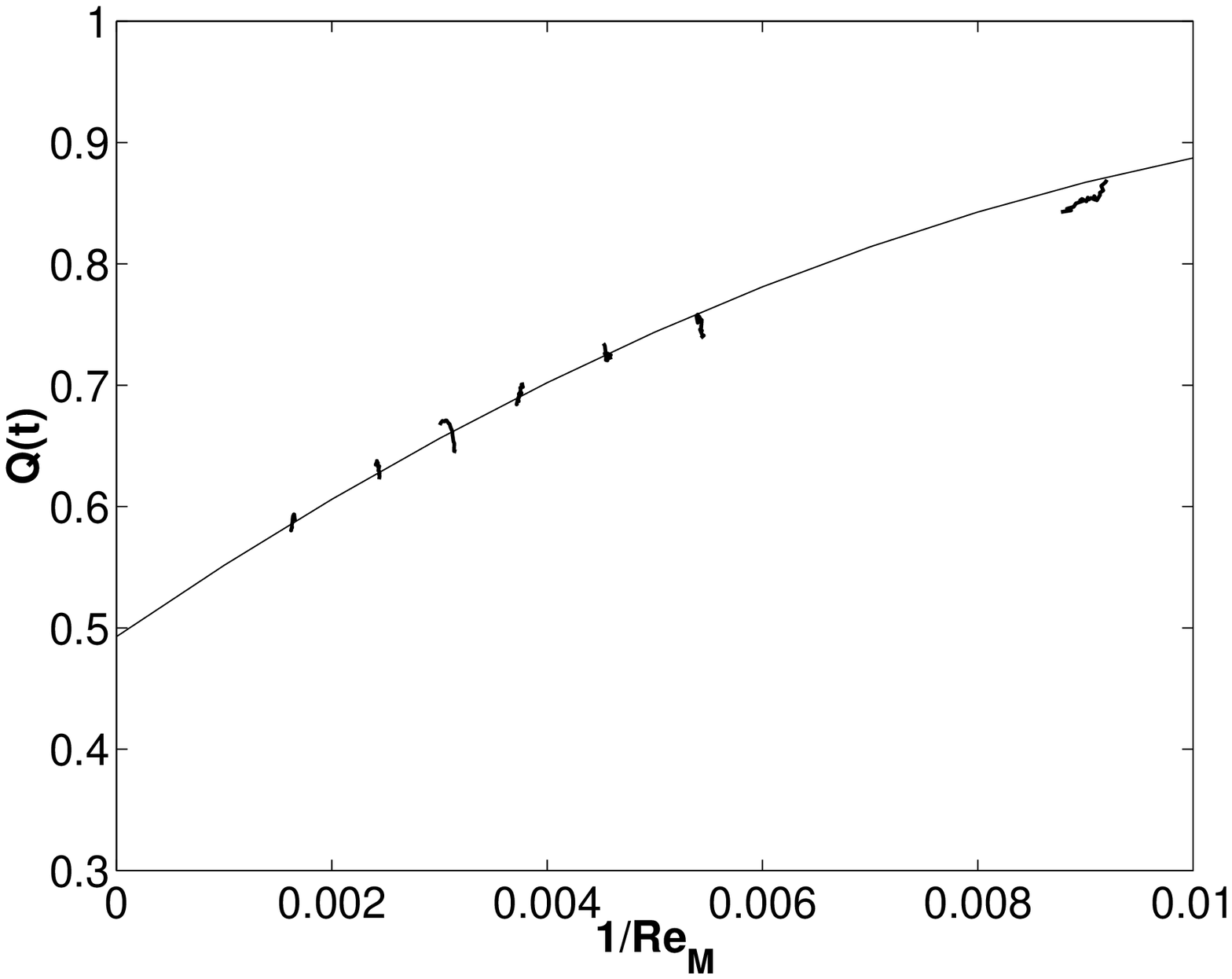}}
}
\caption{(a, left) The quantity $R(t)$, defined by Eq.\ (\ref{RR}) versus 
the quantity $s(t) = (\xi_{\rm diff}/\xi_H)^2$. 
The length scales $\xi_H$ and $\xi_{\rm diff}$ are defined  
in Eq.\  (\ref{length_H}) and the subsequent  
text. (b, right) Energy decay exponent $Q$, defined in Eq.\ (\ref{RR}), plotted against 
the inverse magnetic Reynolds number $Re^{-1}_{\rm M}$,
calculated using the helicity scale $\xi_{\rm H}$. 
The fit is a quadratic least squares fit using the final values.
}
\label{s_dot}
\end{figure}

To characterize the decay laws  we define
\ben
  Q(t) = -t{\dot E}_{\rm M}/E_{\rm M}, \qquad R(t) = -t\dot{H}/2{H}
  \label{RR}
.\een
In Fig.~\ref{s_dot}a we have plotted $R(t)$ versus the quantity
$s(t) = (\xi_{\rm diff}/\xi_H)^2$ for several runs with different initial 
conditions. 
%*****
%*****
This figure tells us that the value 
of $R$ is approximately independent of time which confirms the power law decay
of $H$. 
Fig.~\ref{s_dot}a also indicates that the quantity $s$ is also approximately 
independent of time, hence reinforcing the relation   
$\xi_H \sim t^{0.5}$. 

Taking the relative helicity to be constant, it follows from the power-law
behavior of $H$ that Eq.~(\ref{energy_dec}) is the energy decay law.  In the
limit of exact conservation of magnetic helicity, $s\rightarrow 0$, the magnetic
energy must decay according to the power law $E_{\rm M} \sim t^{-0.5}$.

The parameter $s$ has interesting physical significance.  Note that if $\xi_{\rm
H} \simeq v t$, where $v$ is the RMS velocity, (i.e.\ if the eddy turn-over time
is $t$) then $s \simeq (2\pi)^{2} / Re_{\rm M}$, where $Re_{\rm M}$ is the
magnetic Reynolds number evaluated using the helicity scale $\xi_{\rm H}$.  We
have measured the ratio $f=vt/\xi_{\rm H}$ and $Re^{-1}_{\rm M}$ for all runs,
and find that they are both approximately constant, and are linearly related.
This means there should be a linear relation between $s$ and $Re^{-1}_{\rm M}$,
and hence a quadratic relation between $Q(t)$, the energy decay exponent defined
in Eq.\ (\ref{RR}), and $Re^{-1}_{\rm M}$.  Fig.~\ref{s_dot}b, showing $Q$ and
$Re^{-1}_{\rm M}$, confirms that this is indeed the case, with asymptote at
large $Re_{\rm M}$ consistent with $Q=0.5$.  Finally, assuming only that the
magnetic energy power spectrum is self-similar, and that Ohmic dissipation, if
not dominant, always contributes a constant fraction to the energy loss, one can
show\cite{Christensson:2002xu} that the characteristic length scale of the field
$\xi$ must scale as $t^{0.5}$, thus justifying the assumption made at the start
of this section.

\section{Discussion and conclusions}

To summarize, we have studied the evolution of decaying 3D MHD turbulence involving 
maximally helical magnetic fields. For finite magnetic diffusivity there emerges  
an important quantity $s = (\xi_{\rm diff}/\xi_{\rm H})^2$, where $\xi_{\rm H}$ 
is the helicity scale defined in Eq.\ (\ref{length_H}), and $\xi_{\rm diff}$ is the diffusion 
scale.  We find $\xi_{\rm H} \simeq vt$, where $v$ is the RMS velocity, and hence that 
$s \propto Re^{-1}_{\rm M}$, the magnetic Reynolds number evaluated using the 
helicity scale.  
The magnetic field coherence length (which can be equally well expressed 
as the integral, helicity or relative helicity scales) goes as
$\xi \sim t^{0.5}$, magnetic helicity $H_{\rm M} \sim t^{-2s}$ and
magnetic energy $E_{\rm M} \sim t^{-0.5 - 2s}$.  
A corollary is that $Re_{\rm M}$ is constant once the system has 
reached self-similarity.
Furthermore, we can extrapolate to the limit of
very large magnetic Reynolds numbers, useful for example in the early
Universe, to find $H$ constant and $E_{\rm M} \sim t^{-0.5}$.

%\section*{Acknowledgments}

%This work was conducted on the Cray T3E and SGI Origin platforms using COSMOS 
%Consortium facilities, funded by HEFCE, PPARC and SGI. We also acknowledge
%computing support from the Sussex High Performance Computing Initiative. 
%MH thanks for NORDITA for hospitality, and MC the Astronomy Centre at the University 
%of Sussex.


\begin{thebibliography}{99}

\bibitem{Zel88}Ya.B. Zeldovich, A.A. Ruzmaikin and D.D. Sokoloff, 
{\em Magnetic Fields in Astrophysics} (Gordon \& Breach, New York, 1983); 
A.A. Ruzmaikin, A.A.Shukurov and D.D. Sokoloff, 
{\em Magnetic Fields in Galaxies} (Kluwer, Dordrecht, 1988).

\bibitem{Kro94} P.P. Kronberg, Rep. Prog. Phys. {\bf 57} (1994) 325.

%\cite{Gnedin:2000ax}
\bibitem{GneFerZwe00}
N.~Y.~Gnedin, A.~Ferrara and E.~G.~Zweibel,
%``Generation of the Primordial Magnetic Fields during Cosmological Reionization,''
Astrophys.\ J.\  {\bf 539}, 505 (2000)
[arXiv:astro-ph/0001066].
%%CITATION = ASTRO-PH 0001066;%%

%\cite{Grasso:2000wj}
\bibitem{Grasso:2000wj}
D.~Grasso and H.~R.~Rubinstein,
%``Magnetic fields in the early universe,''
Phys.\ Rept.\  {\bf 348}, 163 (2001)
[arXiv:astro-ph/0009061].
%%CITATION = ASTRO-PH 0009061;%%

\bibitem{BisMul}  D. Biskamp and W.C. M\"uller, 
	Phys. Rev. Lett.{\bf 83}, 2195 (1999);
W.C. M\"uller and D. Biskamp,
	Phys. Rev. Lett.{\bf 84}, 475 (2000).

\bibitem{ChrHinBra01} 
M.~Christensson, M.~Hindmarsh and A.~Brandenburg,
%``Inverse cascade in decaying 3D magnetohydrodynamic turbulence,''
Phys.\ Rev.\ E {\bf 64} (2001) 056405
[arXiv:astro-ph/0011321].

%\cite{Christensson:2002xu}
\bibitem{Christensson:2002xu}
M.~Christensson, 
M.~Hindmarsh and A.~Brandenburg,
%``Scaling laws in decaying helical 3D magnetohydrodynamic turbulence,''
arXiv:astro-ph/0209119.
%%CITATION = ASTRO-PH 0209119;%%

\bibitem{HelGen}
%\cite{Joyce:1997uy}
%\bibitem{Joyce:1997uy}
M.~Joyce and M.~E.~Shaposhnikov,
%``Primordial magnetic fields, right electrons, and the Abelian anomaly,''
Phys.\ Rev.\ Lett.\  {\bf 79}, 1193 (1997)
[arXiv:astro-ph/9703005];
%%CITATION = ASTRO-PH 9703005;%%
%\cite{Vachaspati:2001nb}
%\bibitem{Vachaspati:2001nb}
T.~Vachaspati,
%``Estimate of the primordial magnetic field helicity,''
Phys.\ Rev.\ Lett.\  {\bf 87}, 251302 (2001)
[arXiv:astro-ph/0101261].
%%CITATION = ASTRO-PH 0101261;%%

\bibitem{PouFriLeo76} A. Pouquet, U. Frisch, and J. Leorat,
        J. Fluid. Mech. {\bf 77}, 321 (1976).

\bibitem{Ole97} 
%\bibitem{Olesen:1996ts}
P.~Olesen,
%``On Inverse Cascades in Astrophysics,''
Phys.\ Lett.\ B {\bf 398} (1997) 321
[arXiv:astro-ph/9610154].

%\cite{Son:1999my}
\bibitem{Son:1999my}
D.~T.~Son,
%``Magnetohydrodynamics of the early universe and the evolution of  primordial m
%agnetic fields,''
Phys.\ Rev.\ D {\bf 59} (1999) 063008
[arXiv:hep-ph/9803412].
%%CITATION = HEP-PH 9803412;%%

%\cite{Field:2000hi}
\bibitem{Field:2000hi}
G.~B.~Field and S.~M.~Carroll,
%``Cosmological magnetic fields from primordial helicity,''
Phys.\ Rev.\ D {\bf 62} (2000) 103008
[arXiv:astro-ph/9811206].
%%CITATION = ASTRO-PH 9811206;%%

%\cite{Shiromizu:1998bc}
\bibitem{Shiromizu:1998bc}
T.~Shiromizu,
%``Inverse Cascade of Primordial Magnetic Field in MHD Turbulence,''
Phys.\ Lett.\ B {\bf 443} (1998) 127
[arXiv:astro-ph/9810339].
%%CITATION = ASTRO-PH 9810339;%%

\bibitem{Bis93} D. Biskamp, {\em Nonlinear Magnetohydrodynamics} 
(Cambridge Univ.\ Press, Cambridge, 1993).

\bibitem{BraEnqOle96} 
A.~Brandenburg, K.~Enqvist and P.~Olesen,
%``Large-scale magnetic fields from hydromagnetic turbulence in the very early universe,''
Phys.\ Rev.\ D {\bf 54} (1996) 1291
[arXiv:astro-ph/9602031].

%\cite{Subramanian:1998gi}
\bibitem{Subramanian:1998gi}
K.~Subramanian and J.~D.~Barrow,
%``Magnetohydrodynamics in the early universe and the damping of nonlinear Alfven
% waves,''
Phys.\ Rev.\ D {\bf 58} (1998) 083502
[arXiv:astro-ph/9712083].
%%CITATION = ASTRO-PH 9712083;%%


\bibitem{Bra01}A. Brandenburg, 
Astrophys. J. {\bf 550}, 824 (2001).

%------------------


\end{thebibliography}
\end{document}